\documentstyle[aps,prb,multicol,epsf]{revtex}

\begin{document}

\title{
Hubbard chains network on corner-sharing tetrahedra: 
origin of the heavy fermion state in ${\rm LiV_2O_4}$
}

\author{Satoshi Fujimoto}
\address{
Department of Physics,
Kyoto University, Kyoto 606-8502, Japan
}

\date{\today}
\maketitle
\begin{abstract}
We investigate the Hubbard chains network model defined 
on corner-sharing tetrahedra (the pyrochlore lattice) which is
a possible microscopic model for the heavy fermion state of
${\rm LiV_2O_4}$.
Based upon this model, we can explain transport, magnetic, and thermodynamic
properties of ${\rm LiV_2O_4}$. 
We calculate the spin susceptibility, 
and the specific heat coefficient, 
exploiting the Bethe ansatz exact solution of the 1D Hubbard model
and bosonization method.
The results are quite consistent with experimental observations.
We obtain the large specific heat coefficient 
$\gamma\sim 222 {\rm mJ/mol K^2}$.
\end{abstract}


\begin{multicols}{2}
\section{Introduction}

Recently, it is discovered that a heavy fermion (HF) state is realized in
${\rm LiV_2O_4}$, a transition-metal oxide 
with a cubic spinel structure.\cite{kondo,chm,kondo2,fujiwara,takagi,lee}
It has the largest specific heat coefficient among $d$-electron systems,
$\gamma\sim 420 {\rm mJ/mol K^2}$.\cite{kondo}
The origin of the HF behavior without almost localized 
$f$-electrons is still an important open issue.
Ab initio band calculations suggest that geometrical
frustration of the spinel structure in which 
V sites form corner-sharing tetrahedra network(so-called pyrochlore lattice)
may be important.\cite{band2,band3,band1}
Actually, it is known that 
some geometrically frustrated electron systems in the vicinity of the
Mott transition shows a HF behavior.\cite{shiga,sato,fuji}
However, since the electron density of ${\rm LiV_2O_4}$ is quarter-filling, 
the above idea may not be directly applicable to this system.
Nevertheless, it is expected that the geometrical frustration plays
some crucial role.  
On the other hand, Anisimov et al. proposed that
the $t_{2g}$ band of V sites is splitted into localized and conduction parts
by trigonal crystal field, and thus the system is simulated by
the Kondo lattice model.\cite{ani}

\begin{figure}
{FIG. 1. Some portions of one-dimensional-like bands 
formed by $t_{2g}$ orbital on the pyrochlore lattice.}
\end{figure}

Here, we would like to propose another microscopic model 
for the HF state from a quite different point of view.
We pay attention to the fact that as was shown by band calculations, 
electrons near the Fermi surface consist mainly of the $t_{2g}$ orbitals
of V ion which is quarter-filled, and that the hybridization with $p$-electrons
of oxygen is small.\cite{band2,band3} 
We will neglect trigonal field splitting,
since it is smaller than the band width by $\sim 1/10$.\cite{band2}
Then we can see that each $t_{2g}$ orbital on the corners of 
the pyrochlore lattice forms one-dimensional (1D) like bands along each edge
of tetrahedra (see FIG.1).
It is expected that the hybridization between these 1D bands 
is much suppressed by the geometrical configuration.
In other words, we can say that to reduce the geometrical frustration
the electronic structure maintains the 1D-like character.
Actually, the band structure obtained from this consideration
resembles qualitatively that of ab initio calculations.\cite{band2,band3}
This proposal is also supported by the following experimental observations.
(i) Recent neutron scattering measurements shows that there exists
antiferromagnetic spin fluctuation with a staggered wave vector 
$Q\sim0.84a^{*}$.\cite{lee}
This $Q$ vector is close to that of the quarter-filled
1D Hubbard model. ($Q_{\rm 1D}=2\pi/(4d_{V-V})\sim 0.71a^{*}$.
$d_{V-V}$ is the distance between the nearest neighbor V sites. See ref. 6)
(ii) As temperature is raised, the resistivity increases 
monotonically like $\sim T$ for $T>T^{*}$,
and is not saturated in contrast to $f$-electron-based 
HF systems.\cite{takagi} Here $T^{*}$ is the characteristic temperature
which is analogous to the Kondo temperature of $f$-electron-based
HF systems. 
The behavior is easily understood, 
if we identify $T^{*}$ with a dimensional crossover temperature.
Namely, $T^{*}$ is regarded as
a crossover temperature from 1D to 3D.
For $T>T^{*}$, the 1D-like electronic structure 
gives rise $T$-linear resistivity.\cite{gia}
(iii) Moreover, the Hall coefficient measured by Urano et al. is much small
for $T>T^{*}$.\cite{takagi}
Since there exist several Fermi surfaces in this system,
it is rather difficult to calculate the Hall coefficient
from a microscopic model.  
However, this experimental fact does not contradict with
the interpretation that for  $T>T^{*}$ the system
has 1D-like character. 
We show schematically the temperature dependence of the Hall
coefficient and the resistivity of our model in FIG.2.
It is noted that these properties are not changed qualitatively
even if the sample is not a single crystal, as in the case
of some experimental situations.

Based upon these considerations, we construct the network of
quarter-filled Hubbard chains in directions $(1,\pm 1,0)$, 
$(1,0,\pm 1)$, and $(0,1,\pm 1)$, 
which is defined on the corner-sharing tetrahedra.
We assume that the chains are coupled weakly with each other
at the corners of tetrahedra.
Because of 1D-like structure, electron correlation effect
is much enhanced, leading to the large specific heat coefficient. 
In this paper, we demonstrate this scenario by microscopic calculation.
We show that the temperature dependence of the spin susceptibility 
calculated from our model is 
consistent with experimental observations, and that
the specific heat coefficient is enhanced like 
$\gamma\sim 222 {\rm mJ/mol K^2}$.
A similar idea has been recently proposed by Fulde et al. who
considered a network of 1D Heisenberg spin chains with spins $1$ 
and $1/2$.\cite{fulde} 
However since ${\rm LiV_2O_4}$ is metallic, we believe that
our model is more appropriate to discuss the HF state of this system.  

\begin{figure}
\centerline{\epsfxsize=5.5cm \epsfbox{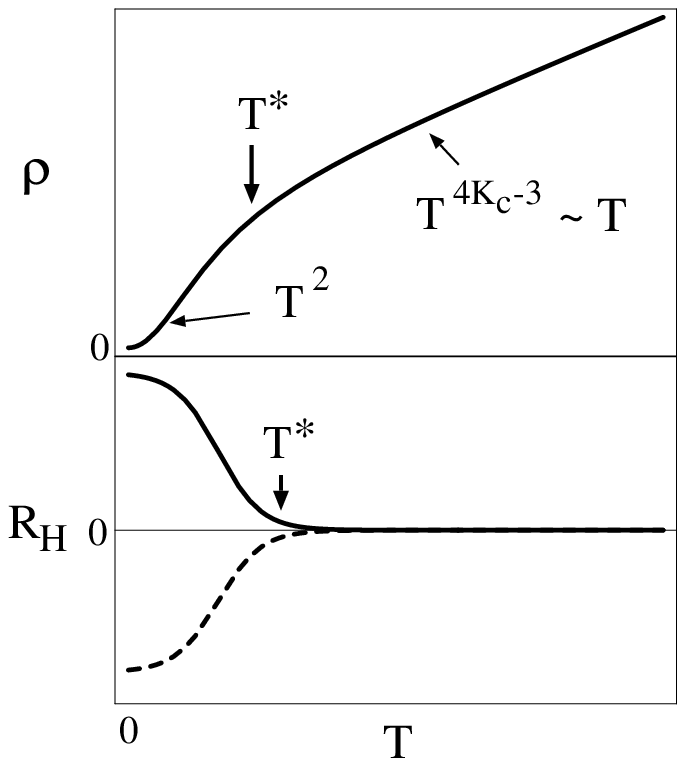}}
{FIG. 2. A schematic view for the temperature dependence of the resistivity
$\rho$ and the Hall coefficient $R_H$ of the Hubbard chains network model.
In high $T$ regions, the Luttinger liquid parameter $K_c\sim 1$.
The sign of $R_H$ for $T<T^{*}$ depends on the curvature of 
the 3D Fermi surface formed
in the low energy scale.}
\end{figure}
 
\section{Hubbard chains network model}

Based on the consideration presented in the introduction,
we construct the Hamiltonian of our model in the presence of 
a magnetic field $h(x_i)$,
\begin{eqnarray}
H&=&\sum_{\alpha=1}^6 H_{\rm 1D}^{(\alpha)}+\sum_{\alpha\beta,\sigma,i}
V_{\alpha\beta}c^{\dagger}_{\alpha\sigma i}c_{\beta\sigma i} 
-g\mu_{\rm B}\sum_i S^z_{\alpha i}h(x_i), 
\end{eqnarray}
where $H_{\rm 1D}^{(\alpha)}$ is the Hamiltonian of the 1D Hubbard model,
\begin{eqnarray}
H_{\rm 1D}^{(\alpha)}&=&-t\sum_{\sigma,i}c^{\dagger}_{\alpha\sigma i}
c_{\alpha\sigma i+1}+h.c.+U\sum_{i}n_{\alpha\uparrow i}n_{\alpha\downarrow i}.
\end{eqnarray}
$c_{\alpha\sigma i}$ ($c^{\dagger}_{\alpha\sigma i}$) 
is an annihilation (creation)
operator for electrons on $\alpha$ chains, and
$ S^z_{\alpha i}=c^{\dagger}_{\alpha\sigma i}(\sigma_z/2)_{\sigma\sigma'}
c_{\alpha\sigma' i}$.
$H^{(1)}_{\rm 1D}$, $H^{(2)}_{\rm 1D}$, 
$H^{(3)}_{\rm 1D}$,...$H^{(6)}_{\rm 1D}$ 
correspond to the chains in the $(1,1,0)$, $(0,1,1)$,
$(1,0,1)$, $(0,1,-1)$, $(1,0,-1)$, and $(1,-1,0)$ directions, respectively.
Because of $t_{2g}$ symmetry of $d$-electrons,
each chain can not hybridize directly at the same points.
However, in ${\rm LiV_2O_4}$, it is possible that
the hybridization between different chains
realizes through $p$-electrons of oxygen.
Actually the hybridization between the chains should be non-local.
This non-locality makes the model quite complicated.
Thus for simplicity, we assume that each Hubbard chains 
hybridizes at each corner of a tetrahedron.
The hybridization parameter $V_{\alpha\beta}$ is given by,
\begin{eqnarray}
\hat{V}=V_{\alpha\beta}= 
\left(
\begin{array}{cccccc}
0&V&V&V&V&0 \\
V&0&V&0&V&V \\
V&V&0&V&0&V \\
V&0&V&0&V&V \\
V&V&0&V&0&V \\
0&V&V&V&V&0 
\end{array}
\right).
\end{eqnarray}
Then the hybridization gives rise the doubling of 1D bands generating total
12 bands because of the
geometrical structure of a tetrahedron.
In spite of this simplification of the hybridization parameter,
the band structure of this network model is roughly
similar to that obtained by first principles band 
calculations.\cite{band2,band3} 
Thus we believe that our model captures the important features
of ${\rm LiV_2O_4}$.
Comparing the band structure with that in refs.\cite{band2} and \cite{band3}, 
we choose $t=0.25$ eV.
We use this parameter in the following calculations. 

The issue of dimensional crossover from 1D to 3D has been studied extensively
so far.\cite{sca,wen,boi}
The basic idea of these previous works is to expand the free energy
in terms of inter-chain couplings, and to apply Landau-Ginzburg type
arguments.
To avoid confusion, we would like to stress that 
our system is not an assembly of parallel chains as was considered 
in the previous studies, but a network composed of chains
aligned in six different directions reflecting the topology
of tetrahedra.
However we can apply the basic method of the dimensional crossover problem
to our model.
Following Boies et al.\cite{boi},
we use the Hubbard-Stratonovich transformation
$ V_{\alpha\beta}c^{\dagger}_{\alpha\sigma i}c_{\beta\sigma i} 
\rightarrow
V_{\alpha\beta}c^{\dagger}_{\alpha\sigma i}\psi_{\beta\sigma}(x_i)
+V_{\alpha\beta}\psi^{\dagger}_{\alpha\sigma}(x_i)c_{\beta\sigma i}
+V_{\alpha\beta}\psi^{\dagger}_{\alpha\sigma}(x_i)\psi_{\beta\sigma}(x_i)$. 
Averaging over $c_{\alpha\sigma i}$ and $c_{\alpha\sigma i}$ with respect to
the action of the 1D Hubbard model, we carry out the cumulant expansion
in $V_{\alpha\beta}$ and $h$.
Then, we have the partition function,
\begin{eqnarray}
Z=Z_{\rm 1D}\int {\cal D}\psi {\cal D}\psi^{\dagger}
e^{-S(\psi,\psi^{\dagger})}, \label{partf}
\end{eqnarray}
where $Z_{\rm 1D}$ is the partition function of the 1D Hubbard model.
The effective action is
given by,
\begin{eqnarray}
S(\psi,\psi^{\dagger})&=&S^{(1)}+\sum_{n=2}^{\infty}S^{(n)} 
+\sum_q\langle S^z(q)S^z(-q)\rangle^{(\rm 1D)}h_q^2 \nonumber \\
&+&\sum V_{\alpha\beta_1}V_{\beta_2\alpha}\langle c_{\alpha\sigma_1 i_1}
c_{\alpha\sigma_2 i_2}^{\dagger}S^z_{\alpha i_3}\rangle^{(\rm 1D)} \nonumber \\
&\times&\psi^{\dagger}_{\beta_1\sigma_1}(x_1)\psi_{\beta_2\sigma_2}(x_2)h(x_3),
\label{effac} \\
S^{(1)}&=&\int dx_1dx_2\psi^{\dagger}_{\sigma}(x_1)
[\hat{V}\delta(x_1-x_2) \nonumber \\
&&-\hat{V}^2\hat{G}^{(1)}(x_1-x_2)]\psi_{\sigma}(x_2), \label{eff1}\\ 
S^{(n)}&=&\frac{1}{n!}\sum G^{(2n)}_{c \alpha}(x_1,x_2,...,x_{2n})
V_{\alpha\beta_1}V_{\beta_2\alpha}...V_{\beta_{2n}\alpha} \nonumber \\
&&\times
\psi^{\dagger}_{\beta_1\sigma_1}(x_1)...\psi_{\beta_{2n}\sigma_{2n}}(x_{2n}),
\label{eff2}
\end{eqnarray}
\noindent
where $x_n=(\mbox{\boldmath $x$}_n,t_n)$, and 
$\langle\cdot\cdot\cdot\rangle^{(\rm 1D)}$ is a correlation function
of the 1D Hubbard model.
$\{\hat{G}^{(1)}(x)\}_{\alpha\beta}=G^{(1)}_{\alpha}(x)\delta_{\alpha\beta}$ 
with $G^{(1)}_{\alpha}(x)$ the single-particle Green's function of 
$\alpha$ chain, and 
$G^{(2n)}_{c \alpha}(x_1,...,x_{2n})$ is the connected 
$2n$-particle correlation function of $\alpha$ chain, 
$\langle c_{ 1}c^{\dagger}_{ 2}...c_{2n-1}c^{\dagger}_{2n}
\rangle_c^{(\rm 1D)}$. $\hat{V}$ is a matrix with elements 
$V_{\alpha\beta}$. 

In the presence of the hybridization $V$, the quasiparticle weight,
which is the hallmark of the Fermi liquid state,
develops as the temperature is lowered.\cite{boi}
To see this, it is sufficient to consider only up to $S^{(1)}$ term of 
eq.(\ref{effac}).
The pole of the single particle Green's function
$\hat{G}(k,\varepsilon)=\hat{G}^{(1)}(k,\varepsilon)
[1-\hat{V}\hat{G}^{(1)}(k,\varepsilon)]^{-1}$ 
determines the quasiparticle energy spectrum
$E_k\sim [2(1-\theta)v_cv_s/(v_c+v_s)]k\equiv v_Fk$.
Here $v_c$ is the velocity of holon, and $v_s$ is that of spinon.
The quasiparticle weight is $z\sim  
2\sqrt{v_cv_s}(\Delta kv_c/E_F)^{\theta}/(v_c+v_s)$,
with $\Delta k=(4Vv_c^{\theta}(v_cv_s)^{-1/2}E_F^{-\theta})^{1/(1-\theta)}$,
$\theta=(K_c+1/K_c-2)/4$, $K_c$ the Luttinger liquid parameter
of the charge sector, 
and $E_F$ is the Fermi energy 
of the 1D Hubbard model.
$S^{(n)}$ ($n\geq 2$) terms are relevant perturbations to
this Fermi liquid state.
As long as spontaneous symmetry breaking does not occur, the Fermi liquid
state is stable.
At sufficiently high temperatures, we can treat $S^{(n)}$ ($n\geq 2$) terms
as small perturbations.
In the following, we take into account only $S^{(2)}$ term.
At low temperatures, effects of 
higher order terms $S^{(n)}$ ($n\geq 3$) are not negligible.
To estimate the temperature range in which our approach is valid,
we apply simple scaling argument to the action (\ref{effac}).
The scaling equation for the effective coupling of $S^{(2)}$, which we denote
$g_4$, is given by, 
\begin{eqnarray}
\frac{d g_4}{d {\rm ln}(\frac{T_F}{T})}=(1-K_c)g_4.
\end{eqnarray}
Then, $g_4$ grows to the order of unity at $T\sim (V/4t)^{4/(1-K_c)}T_F$,
where $T_F\sim 4t$, the band width of the 1D Hubbard model.
Thus our approach is applicable for $T > T_0\equiv (V/4t)^{4/(1-K_c)}T_F$.
For $T<T_0$, our approximation will be less accurate quantitatively.
As we will see later, $T_0$ is sufficiently small for the parameters
we use.

Using the bosonization rule and the operator product expansion
for the $U(1)$ Gaussian model and 
the level-1 $SU(2)$ Wess-Zumino-Witten model, we can show that the leading
term of $S^{(2)}$ is written as,\cite{hal,kad,ope1}
\end{multicols}
\begin{eqnarray}
&&S^{(2)}\sim \frac{1}{2}\sum_{\alpha\beta_1...\beta_4}\sum_{k,k',q}
V_{\alpha\beta_1}V_{\beta_2\alpha}
...V_{\beta_{4}\alpha}[\frac{1}{4v_c^2}
\langle\rho_{\alpha}(q)\rho_{\alpha}(-q)
\rangle^{({\rm 1D})}
\sum_{\sigma\sigma'}\psi_{\beta_1\sigma}^{\dagger}(k'+q)
\psi_{\beta_2\sigma }(k')
\psi_{\beta_3\sigma'}^{\dagger}(k-q)\psi_{\beta_4\sigma'}(k) \nonumber \\
&&+\frac{1}{2v_s^2}\langle S^{+}_{\alpha}(q)S^{-}_{\alpha}(-q)
\rangle^{({\rm 1D})}
\sum_{\sigma_1\sigma_2\sigma_3\sigma_4}
\psi_{\beta_1\sigma_1}^{\dagger}(k'+q)
\mbox{\boldmath $\sigma$}_{\sigma_1\sigma_2}
\psi_{\beta_2\sigma_2}(k')
\psi_{\beta_3\sigma_3}^{\dagger}(k-q)
\mbox{\boldmath $\sigma$}_{\sigma_1\sigma_2}
\psi_{\beta_4\sigma_4}(k)].
\end{eqnarray}
\begin{multicols}{2} 
\noindent
Here $\rho(q)=\sum_{\sigma,k}c^{\dagger}_{\sigma,k+q}c_{\sigma, k}$, 
$S^{+}_{\alpha}(q)=\sum_{k}c^{\dagger}_{\uparrow,k+q}c_{\downarrow,k}$, etc.
In a similar manner, we can rewrite the last term of eq.(\ref{effac}) as,
\begin{eqnarray}
&&\frac{1}{v_s}\sum_{\alpha\beta_1\beta_2}
V_{\alpha\beta_1}V_{\beta_2\alpha}[\sum_{q\sim 0}\sum_{\omega} 
\chi^{\rm 1D}_u(q,\omega)\psi^{\dagger}_{\beta_1 \sigma_1, k+q}
\sigma^z_{\sigma_1\sigma_2}
\psi_{\beta_2 \sigma_2, k}h_q \nonumber \\
&&+\sum_{q\sim Q}\sum_{\omega} 
\chi^{\rm 1D}_s(q,\omega)\psi^{\dagger}_{\beta_1 \sigma_1, k+q}
\sigma^z_{\sigma_1\sigma_2}
\psi_{\beta_2 \sigma_2, k}h_q]. 
\end{eqnarray}
Here $\chi^{\rm 1D}_u(q\sim 0,\omega)$ and 
$\chi^{\rm 1D}_s(q\sim Q_0,\omega)$ are 
the uniform and staggered parts 
of the spin susceptibility, respectively, for the 1D Hubbard model.
$Q_0=\pi/2$ in the quarter-filling case.

\section{Magnetic properties}

In this section we consider the magnetic properties of the system.
Applying random phase approximation, 
we compute the spin susceptibility, 
\begin{eqnarray}
\chi(q)&=&-T\delta^2\ln Z/\delta h_q^2 \nonumber \\
&=&
\sum_{\alpha\beta}
\{\hat{\chi}^{1D}(q)(1-\hat{\Gamma}(q)\hat{\chi}^{1D}(q))^{-1}\}_{\alpha\beta},
\label{spinsus}
\end{eqnarray} 
where $\{\hat{\chi}^{\rm 1D}(q)\}_{\alpha\beta}
=\chi^{\rm 1D}_{\alpha}(q_{\alpha})\delta_{\alpha\beta}$ with
$q_1=q_2=q_x+q_y$, $q_3=q_4=q_y+q_z$, $q_5=q_6=q_x+q_z$,
$q_7=q_8=q_y-q_z$, $q_9=q_{10}=q_x-q_z$, and $q_{11}=q_{12}=q_x-q_y$.
$\{\hat{\Gamma}(q)\}_{\alpha\beta}=\frac{1}{v_s^2}
\sum_k \tilde{G}^{0}_{\alpha\beta}(k+q)\tilde{G}^{0}_{\alpha\beta}(k)$,
with $\tilde{G}^{0}_{\alpha\beta}=\{\hat{V}\hat{G}^{0}
\hat{V}\}_{\alpha\beta}$, and
$\hat{G}^{0}=[\hat{V}-\hat{V}^2\hat{G}^{(1)}]^{-1}$.
It is noted that in the denominator of eq.(\ref{spinsus})
the single-particle Green's function $\hat{G}^0$ appears.
Eq.(\ref{spinsus}) implies that as the quasiparticle weight develops,
the spin-spin correlation between Hubbard chains 
which is mediated by two particle hopping
is enhanced. This point is important for the following arguments.
Although we know the low energy expression of $G^{(1)}(x,t)$, 
it is difficult
to carry out the Fourier transform analytically.\cite{bethe}
For simplicity, we approximate the anomalous exponent of $G^{(1)}(x,t)$ as 
$\theta\approx 0$. 
Actually $\theta$ is sufficiently small 
for intermediate strength of $U$, {\it e.g.}
$\theta=0.059$ ($K_c=0.618$) for $U/4t=2$, and $\theta=0.03$ 
($K_c=0.71$) for $U/4t=1$.
Then the Fourier transform of $G^{(1)}(x,t)$ is given by,
\begin{eqnarray}
G^{(1)R}_{L}(k,\varepsilon)&=&\frac{-2iA\sqrt{v_cv_s}}{(2\pi)^2T(v_c-v_s)}
B(\frac{1}{4}-i\frac{v_s(\varepsilon-kv_c)}
{2\pi(v_c-v_s)T},\frac{1}{2}) \nonumber \\
&&\times B(\frac{1}{4}
-i\frac{v_c(\varepsilon-kv_s)}{2\pi(v_c-v_s)T},
\frac{1}{2}),
\label{gf}
\end{eqnarray}
where $B(x,y)$ is the beta function, and
$A$ is a non-universal pre-factor which depends on model parameters.
Unfortunately we do not know the exact value of $A$.
In the following, to avoid overestimating electron correlation effects,
we use the non-interacting value of $A$.
We now calculate the uniform susceptibility $\chi(q=0)$ from
eqs.(\ref{spinsus}) and (\ref{gf}).
In the numerical calculation, we neglect the off-diagonal components
of $\hat{G}^{0}$ and $\hat{\chi}$, which may give subdominat corrections.
We choose the parameter $4t=1$ eV, and $U=2$ eV, the value suggested from
photo-emission measurements.\cite{fujimori}
Using the Bethe ansatz exact solution, we can obtain $v_c$, $v_s$, 
and $\chi^{\rm 1D}(0)=1/(2\pi v_s)$.\cite{lieb,shiba}
We determine $V$ by fitting the uniform spin susceptibility 
with experimental data.
The result is shown in FIG.3. We set $V/t=0.4506$.
For this parameter, $T_0/T_F\sim 10^{-8}$.
Thus we can apply our results to the wide temperature range including
sufficiently low temperature regions.
The spin susceptibility consists of the Pauli paramagnetic term and
the temperature-independent Van Vleck term.
To fit to experimental data, we set the Van Vleck term
$\chi_{\rm VV}\sim 2.1\times 10^{-3} {\rm emu/mol}$.
It is noted that the Curie-like temperature dependence is not due to
the presence of local moment as in the case of dense Kondo systems,
but caused by the development of 3D-like magnetic correlation between
Hubbard chains.  As temperature decreases, the single-particle 
weight becomes large, leading to the enhancement of 3D correlation mediated
by two particle hopping.

\begin{figure}
\centerline{\epsfxsize=6.5cm \epsfbox{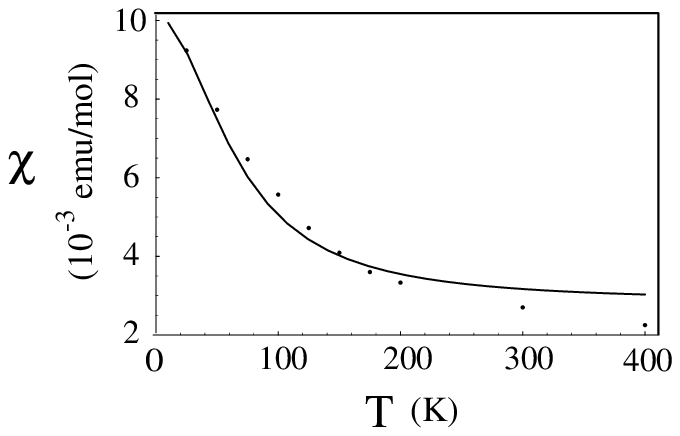}}
{FIG. 3. Calculated result of temperature dependence 
of the spin susceptibility(solid line).
The dots are experimental data quoted from refs.\cite{takagi} 
and \cite{kondo2}.}
\end{figure}

\section{Specific heat coefficient: heavy fermion state}

Let us consider the specific heat coefficient which characterizes
the HF behavior.
According to eq.(\ref{partf}), the free energy is given by
the sum of the 1D part and the Fermi liquid part. 
Thus the specific heat coefficient is
expressed as,
$\gamma=\gamma_{\rm 1D}+\gamma_{\rm FL}$.  
It should be noted that this expression is valid
only around the crossover temperature $T^{*}$,
since in the sufficiently low temperature region the higher order terms
of $S^{(n)}$ which are neglected in our approximation
will cancel the 1D-like contribution, and the conventional Fermi liquid
result must be recovered. 
Thus, $\gamma \sim \gamma_{\rm FL}$ in the low temperature
region where the system is in the 3D Fermi liquid state.

We calculate the Fermi liquid part $\gamma_{\rm FL}$ 
taking into account contributions from 
spin fluctuations.
Since in the 1D Hubbard model the staggered component of the spin
fluctuation is dominant over the uniform one,
the antiferromagnetic spin fluctuation gives leading corrections to
the self-energy.
This is consistent with recent neutron scattering measurements.\cite{lee}
The propagator for the antiferromagnetic spin fluctuation is obtained from
eq.(\ref{spinsus}). 
We calculate numerically
the Fourier transform of the staggered spin susceptibility
$\chi^{\rm 1D}_s(x,t)=e^{iQ_0x}
{\rm Im}\{1/[2\pi^2(x^2-v_c^2(t-i\delta)^2)^{K_c/2}
(x^2-v_s^2(t-i\delta)^2)^{1/2}]\}\theta(t)$,
which is divergent for $q=Q_0$.\cite{bethe}
This divergence is cutoff by the momentum scale
at which the crossover from 1D to 3D system occurs, {\it i.e.}
$q\sim \Delta k$.
Since the non-universal pre-factor of $\chi^{\rm 1D}_s(x,t)$ 
is unknown for finite $U$, we use the pre-factor of
non-interacting systems to simplify the calculation. 
$\chi_{11}(q,\omega)$ has peaks at $\mbox{\boldmath $Q$}
=(Q_0,0,0)$, $(0,Q_0,0)$ 
and $(Q_0/2,Q_0/2,Q_0/2)$. Expanding eq.(\ref{spinsus})
around one of these peaks, we have,
\begin{eqnarray}
\chi_{11}(\mbox{\boldmath $Q$}+\mbox{\boldmath $q$},\omega)
\sim \frac{\chi_0(\mbox{\boldmath $Q$})}{1+\xi_x^2q_x^2
+\xi_y^2q_y^2+\xi_z^2q_z^2-i\omega/\Gamma_q}. \label{spinflu}
\end{eqnarray}
Here $\chi_0(\mbox{\boldmath $Q$})
=\{\hat{\chi}^{\rm 1D}(\mbox{\boldmath $Q$})
(1-\hat{\Gamma}(\mbox{\boldmath $Q$})
\hat{\chi}^{\rm 1D}(\mbox{\boldmath $Q$}))^{-1}]\}_{11}$, and 
\begin{eqnarray}
&&\xi_x^2=\xi_y^2=2[12(b_0\chi^{\rm 1D}_u+b_Q\chi^{\rm 1D}_s)
c_Q\chi^{\rm 1D}_s 
+8b_Qc_0\chi^{\rm 1D}_u\chi^{\rm 1D}_s \nonumber \\
&&+16b_Q^2c_0\chi^{\rm 1D}_u(\chi^{\rm 1D}_s)^2 
+80c_Q\chi^{\rm 1D}_s b_0b_Q\chi^{\rm 1D}_u
\chi^{\rm 1D}_s] \nonumber \\
&&\times 1/(1-A_0), \\
&&\xi_z^2=[16(b_0\chi^{\rm 1D}+b_Q\chi^{\rm 1D}_s)
c_Q\chi^{\rm 1D}_u 
+32b_Qc_0\chi^{\rm 1D}_u\chi^{\rm 1D}_s \nonumber \\
&&+64b_Q^2c_0\chi^{\rm 1D}_u(\chi^{\rm 1D}_s)^2 
+64c_Q\chi^{\rm 1D}_sb_0b_Q\chi^{\rm 1D}_u
\chi^{\rm 1D}_s] \nonumber \\
&&\times 1/(1-A_0), \\
&&A_0=8(b_Q\chi^{\rm 1D}_s+b_0\chi^{\rm 1D}_u)^2 
+16b_0b_Q\chi^{\rm 1D}_u
\chi^{\rm 1D}_s \nonumber \\
&&+64(b_Q\chi^{\rm 1D}_s+b_0\chi^{\rm 1D}_u)b_0b_Q\chi^{\rm 1D}_u
\chi^{\rm 1D}_s,
\end{eqnarray}
with $\chi^{\rm 1D}_u=\chi^{\rm 1D}_u(q=0)$, 
$\chi^{\rm 1D}_s=\chi^{\rm 1D}_s(q=Q_0)$,
$b_0=(zV/v_s)^2/(\pi v_F)$, 
$b_Q=(zV/v_s)^2\ln(4t/v_F\Delta k)/(\pi v_F)$,
$c_0=4(zV/v_s)^2/(3\pi^2 v_F)$, and $c_Q=(zV/v_s)^2/(16\pi v_F\Delta k^2)$.
Using eq.(\ref{spinflu}), we compute the self-energy $\Sigma_{11}$ and
obtain the mass enhancement factor,
\begin{eqnarray}
-\frac{\partial\Sigma_{11}}{\partial\varepsilon}&\sim& 
\frac{192\sqrt{2}V^4\chi_0(\mbox{\boldmath $Q$})}{\pi^2 v_s^2v_F\xi_y\xi_z}
\ln[1+(\frac{\xi_y^2}{2}+\xi_z^2)\pi]\sim 4.33.
\end{eqnarray}
Then, we end up with the specific heat coefficient,
\begin{eqnarray}
\gamma=\biggl(1-\frac{\partial \Sigma}{\partial \varepsilon}
\biggr)
\frac{2\pi}{3v_F}\sim 222 \quad{\rm (mJ/mol K^2)}.
\end{eqnarray}
This is almost consistent with experimentally observed values,
{\it i.e.} $350\sim420 {\rm mJ/mol K^2}$.
Since our approximation underestimates electron correlation effects,
the above result gives a lower bound for $\gamma$.

\section{summary and comments}

In this paper, we have proposed a microscopic scenario for the HF state 
of ${\rm LiV_2O_4}$. It has been shown that magnetic, thermodynamic 
and transport properties of ${\rm LiV_2O_4}$
are well understood in terms of the Hubbard chains network model.
In the high temperature region, the system is in the 1D-like state showing
anomalous transport properties.
As temperature is lowered, the crossover to 3D Fermi liquid state occurs.
The origin of heavy fermion mass is ascribed to enhanced electron
correlation of the 1D-like structure.

The validity of our model completely replies on the assumption
that the system is on the border of the dimensional crossover. 
We need more experimental tests to confirm the validity of our scenario.
Very recently, Li-NMR measurement under ambient pressure has been carried out
by Fujiwara et al.\cite{nmr} 
They found that the spin lattice relaxation rate 
$(T_1T)^{-1}$ increases, as the applied pressure is increased,
showing the enhancement of the spin fluctuation.
This behavior is quite different from that expected for usual
strongly correlated metals like $f$-electron based HF
systems.  
This experimental fact may be easily understood in terms of the dimensional
crossover phenomena described by our model. 

Finally, we make a brief comment on single-particle properties.
It is known that in the 1D Hubbard model at quarter-filling, $v_c$ is much
larger than $v_s$ for large $U$. 
In our case ($U/4t=2$), $v_c=1.9t$ and $v_s=0.6t$.
Thus it is expected that for $T^{*}<T\ll E_F/k_{\rm B}$ 
the single-particle spectrum have
two peaks which correspond to charge and spin degrees of freedom, 
indicating spin-charge separation. 
It may be intriguing to search for spin-charge separation behavior 
in ${\rm LiV_2O_4}$
by photo-emission spectroscope experiment.

\acknowledgements{}
The author would like to thank K. Yamada, H. Tsunetsugu, and N. Fujiwara
for invaluable discussions.
This work was partly supported by a Grant-in-Aid from the Ministry
of Education, Science, and Culture, Japan.




\end{multicols}


\begin{references}
\bibitem{kondo}   S. Kondo, D. C. Johnston, C. A. Swenson, F. Borsa,
A. V. Mahajan, L. L. Miller, T. Gu, A. I. Goldman, M. B. Maple, D. A. Gajewski,
E. J. Freeman, N. R. Dilly, R. P. Dickey, J. Merrin, K. Kojima, G. M. Luke, 
Y. J. Uemura, O. Chmaissem, and J. D. Jorgensen, 
Phys. Rev. Lett. {\bf 78}, 3729 (1997).

\bibitem{chm} O. Chmaissem, J. D. Jorgensen, S. Kondo, and D. C.
Johnston, Phys. Rev. Lett. {\bf 79},
4866 (1997).

\bibitem{kondo2} S. Kondo, D.C. Johnston, and L.L. Miller, 
Phys. Rev. B{\bf 59}, 2609 (1999).

\bibitem{fujiwara} N. Fujiwara, H. Yasuoka, and Y. Ueda, Phys. Rev.
B{\bf 57}, 3539 (1998).

\bibitem{takagi} C. Urano, M. Nohara, S. Kondo, F. Sakai,
H. Takagi, T. Shiraki, and T. Okubo, Phys. Rev. Lett. {\bf 85}, 1052
(2000).

\bibitem{lee} S.-H. Lee, Y. Qiu, C. Broholm, Y. Ueda, and J. J. Rush, 
Phys. Rev. Lett. {\bf 86}, 5554
(2001).

\bibitem{band2} J. Matsuno, A. Fujimori, and L. F. Mattheiss, 
Phys. Rev. B{\bf 60}, 1607 (1999).

\bibitem{band3} D. J. Singh, P. Blaha, K. Schwarz, and I. I. Mazin, 
Phys. Rev. B{\bf 60}, 16359 
(1999).

\bibitem{band1} V. Eyert, K. H. Hock, S. Horn, A. Loidl, and 
P. S. Riseborough, Europhys. Lett. {\bf 46}, 
762 (1999). 

\bibitem{shiga}  M. Shiga, H. Wada, Y. Nakamura, J. Deportes, and
K. R. A. Ziebeck, J. Phys. Soc. Jpn. {\bf 57}, 3141 (1988);
M. Shiga, K. Fujisawa, and H. Wada,{\it ibid} {\bf 62}, 1329 (1993).

\bibitem{sato} S. Yoshii and M. Sato, J. Phys. Soc. Jpn. {\bf 68}, 3034
(1999).

\bibitem{fuji} S. Fujimoto, Phys. Rev. B{\bf 64}, 085102 (2001).

\bibitem{ani} V. I. Anisimov, M. A. Korotin, M. Z\"olfl,
T. Pruschke, and K. Le Hur, and T. M. Rice, Phys. Rev. Lett. {\bf 83},
364 (1999).

\bibitem{gia} T. Giamarchi, Phys. Rev. B{\bf 44}, 2905 (1991).

\bibitem{fulde} P. Fulde, A. N. Yaresko, A. A. Zvyagin, and Y. Grin,
Europhys. Lett., {\bf 54}, 779 (2001).

\bibitem{sca} D. J. Scalapino, Y. Imry, and P. Pincus, Phys. Rev. B{\bf 11},
2042 (1975).

\bibitem{wen} X. G. Wen, Phys. Rev. B{\bf 42}, 6623 (1990).

\bibitem{boi} D. Boies, C. Bourbonnais, and A.-M. S. Tremblay,
Phys. Rev. Lett. {\bf 74}, 968 (1995).

\bibitem{hal} F. D. M. Haldane, J. Phys. C {\bf 14}, 2585 (1981).

\bibitem{kad} L. P. Kadanoff and A. C. Brown, Ann. Phys. {\bf 121},
318 (1979).

\bibitem{ope1} V. Knizhnik and A. Zamolodchikov, Nucl. Phys. B{\bf 247},
83 (1984).

\bibitem{bethe} see, for example, V. J. Emery in {\it Highly 
Conducting One-Dimensional Solids}, ed J. T. DeVreese {\it et al.},
(Plenum, New York, 1979); H. Frahm and V. E. Korepin, Phys. Rev. B{\bf 42},
10553 (1991).  

\bibitem{fujimori} A. Fujimori, K. Kawakami, and N. Tsuda, 
Phys. Rev. B{\bf 38}, 7889 (1988).

\bibitem{lieb} E. H. Lieb and F. Y. Wu, Phys. Rev. Lett. {\bf 20},
1445 (1968).

\bibitem{shiba} H. Shiba, Phys. Rev. B{\bf 6}, 930 (1972).

\bibitem{bethe2} M. Takahashi, Prog. Theor. Phys. {\bf 47}, 69 (1972);
{\it ibid} {\bf 52}, 103 (1974). 

\bibitem{nmr} K. Fujiwara, K. Miyoshi, J. Takeuchi, T. C. Kobayashi, and 
K. Amaya, unpublished.

\end{references}
\end{document}